\documentclass{mpe_report}

\usepackage{psfig,graphicx,epsfig}
\usepackage{color}
\usepackage{amsmath,amssymb,epic,eepic,array}

\begin{document}

\pagenumbering{arabic}
\setcounter{page}{208}

 \renewcommand{\FirstPageOfPaper }{208}\renewcommand{\LastPageOfPaper }{211}

\title{Testing consistency of deconfinement heating of strange stars in
  superbursters and soft X-ray transients}
\author{Morten Stejner\inst{1}}  
\institute{Department of Physics and Astronomy, University of Aarhus,
 Ny Munkegade Bld 520, 8000 Aarhus C, Denmark}
\titlerunning{Deconfinement heating of strange stars}
\maketitle
\begin{abstract}
  Both superbursters and soft X-ray transients probe the thermal
  structure of the crust on compact stars and are sensitive to the
  process of deep crustal heating. It was recently shown that the
  transfer of matter from crust to core in a strange star can heat the
  crust by deconfinement and ignite superbursts provided certain
  constraints on the strange quark matter equation of state are
  fulfilled. Corresponding constraints are derived for soft X-ray
  transients in a simple parameterized model assuming their quiescent
  emission is powered in the same way, and the time
  dependence of this heating mechanism in transient systems is discussed.
\end{abstract}
\section{Introduction}
The possibility that strange stars may carry a small crust of ordinary
nuclear matter separated from the quark matter phase by an electron
filled gap is key to reconciling observed X-ray phenomenology in
accreting binary systems with the strange quark matter hypothesis, because
it allows deconfinement heating near the bottom of the crust to
replace pycnonuclear reactions as the source of deep crustal heating
needed to achieve observed inner temperatures. It was shown by
\cite{Page:2005} that accreting strange stars fulfilling a set of
general constraints on the equation of state may thus achieve
temperatures high enough to ignite superbursts over a broad range of
parameters, and in \cite{Stejner:2006} we recently considered the
consistency of these constraints with observed surface temperatures of
quiescent soft X-ray transients showing that such an approach may
constrain the properties of strange quark matter further. 

Superbursts -- extremely long and energetic type I X-ray bursts
(\cite{Kuulkers:2004} for an observational review) -- are believed to
originate in unstable carbon burning at densities around $10^9 \mbox{
  g cm}^{-3}$ (Cumming et al. 2005). For carbon to ignite at such
densities the temperature must reach $6\times 10^8$ K, and since
superbursters all have similar accretion rates around
$0.1-0.3\,\dot{m}_\mathrm{Edd}$, where $\dot{m}_\mathrm{Edd}$ is the
Eddington accretion rate, this gives an indication of the heat
released as the accreted matter adjusts chemically to the increasing
weight of the layers above. In neutron stars the majority of this heat
($\sim1.4$ MeV/Nucleon) is released by pycnonuclear reactions beyond
the neutron drip density (\cite{Haensel:1990,Haensel:2003,Brown:1998}),
but \cite{Cumming:2005} found that even this may be insufficient to
reach superburst ignition due to the neutrino emissivity associated
with pair breaking and formation in superfluid neutron in the deep
crust. The crust on a strange star is sustained electrostatically
above the quark matter core and is therefore limited to the neutron
drip density -- and probably somewhat below this density due to
tunneling of ions through the Coulomb barrier
(Alcock et al. 1986, Stejner \& Madsen 2005)
%\cite{Alcock:1986,Stejner:2005}. 
This removes the neutrino emissivity
from paired neutrons. Furthermore the deconfinement heating as matter
is transferred from crust to core is potentially very powerful
releasing a binding energy of $Q_\mathrm{SQM} \lesssim 100$ MeV per
nucleon, so accreting strange stars can potentially be hotter than
neutron stars. This would depend however on the thermal conductivity,
$K$, and neutrino emissivity, $\epsilon_\nu$, of the quark matter core
which -- although in principle directly calculable -- depend on poorly
constrained QCD-parameters. Hence superbursts actually constrain the
relationship between the three core parameters, $Q_\mathrm{SQM}$, $K$,
and $\epsilon_\nu$ as shown by \cite{Page:2005}.

Soft X-ray transients are binary systems which go through long periods
(years to decades) of relative quiescence with luminosities around
$10^{32} \mbox{ erg s}^{-1}$ punctuated by short accretion outbursts
giving time averaged accretion rates up to $10^{-10} { M}_\odot \mbox{
  yr}^{-1}$. During quiescence the thermal radiation detected from the
surface of some sources has been shown by Brown et al.(1998) to be
consistent with with deep crustal heating by pycnonuclear reactions in
neutron stars, but as for superbursters this would have to be replaced
with deconfinement heating if the strange quark matter hypothesis holds
true. Given a relationship between core and surface temperature and
taking proper account of cooling by photon emission from the surface
-- which is more important for these cold systems -- soft X-ray
transients therefore probe the relationship between the core
parameters in much the same way as superbursters. 
 
\section{Thermal structure model}
The calculations are performed in a parameterized model, which although
very simple has the virtues of allowing direct comparison and scaling
between different systems and providing a physical insight easily lost
in numerical modelling. This only allows a consistency check however
and a more detailed model would be needed to strongly constrain the
core parameters. The model is briefly summarized, but the
reader is referred to \cite{Stejner:2006} for greater detail.

The thermal structure of the outer crust is discussed in detail in the
literature -- see e.g. \cite{Yakovlev:2004} for a recent account --
but here we use the result from \cite{Gudmundsson:1983} that at
densities greater than $10^{10}\mbox{ g cm}^{-3}$ the crust is
approximately isothermal with the inner temperature $T_\mathrm{i}$
related to the surface temperature $T_\mathrm{S}$ by
$T_\mathrm{i,9}\simeq 0.1 T_\mathrm{S,6}^2$, where
$T_\mathrm{i,9}=T_\mathrm{i} /10^9$ K.

The quark matter core is modelled in a plane parallel approximation with a
heat source from deconfinement at the core surface and a heat sink from
neutrino emission, which to ensure
superbust ignition must take place by slow processes with emissivity
(Page \& Cumming 2005)
%\cite{Page:2005}

\begin{equation}
\epsilon_\nu = Q_\nu T_9^8 \mbox{ erg cm}^{-3}\mbox{ s}^{-1}, \quad
Q_\nu \sim 10^{18}-10^{22}\,.
\end{equation} 
Ignoring cooling by photon emission from the surface -- as
appropriate for superbursters -- this gives an inner temperature
(Page \& Cumming 2005)
%\cite{Page:2005}

\begin{equation}\label{iso}
T_\mathrm{i,9}=0.54\, Q_{\nu,21}^{-1/8}\left(\frac{Q_\mathrm{SQM}}{100\mathrm{
    MeV}}\right)^{1/8}\left(\frac{\dot{m}}{0.3\;\dot{m}_\mathrm{Edd}}\right)^{1/8}R_6^{-1/8}  
\end{equation}
if the thermal conductivity of strange quark matter is high enough for
the core to isothermal. If it is not then 
\begin{equation}\label{niso}
T_\mathrm{i,9}=0.87\, Q_{\nu,21}^{-1/9} K_{20}^{-1/9}\left(\frac{Q_\mathrm{SQM}}{100\mbox{ MeV}}\right)^{2/9}\left(\frac{\dot{m}}{0.3\;\dot{m}_\mathrm{Edd}}\right)^{2/9}
\end{equation}
with $K$ assumed to lie in the range $10^{19}-10^{22}$ cgs. By
comparison, the critical thermal conductivity for the core to be
isothermal is then $K_\mathrm{crit}=4.3\times 10^{21}\mbox{ cgs }
Q_{\nu,21}^{1/8}(Q_\mathrm{SQM}/100\mathrm{
  MeV})^{7/8}(\dot{m}/0.3\;\dot{m}_\mathrm{Edd})^{7/8}$.  These
relations were shown to be reasonable approximations to more detailed
numerical calculations in \cite{Page:2005} and express a relation
between the core parameters, which must be fulfilled in order to reach
superburst ignition conditions (i.e. $T_\mathrm{i,9}=0.7$ with
$\dot{m}=0.1-0.3\,\dot{m}_\mathrm{Edd}$ ).

Including cooling by photon luminosity from the surface -- as
appropriate for colder sources -- deconfinement
heating must balance the neutrino luminosity and the photon luminosity,
$L_\mathrm{DH} = L_\nu + L_\gamma$, which using a pure blackbody and the
relations between the core parameters gained from superbursters and
Eqs. \ref{iso} and \ref{niso} gives 
(Stejner \& Madsen 2006)
%\cite{Stejner:2006}

\begin{equation}
\dot{m}\,Q_\mathrm{SQM}=1.8\times 10^{-7}
  \dot{m}_\mathrm{SB}\,Q_\mathrm{SQM} T_\mathrm{S,6}^{16}+\sigma T_\mathrm{S}^4\label{isocombined}
\end{equation}
for an isothermal core and
\begin{equation}
\dot{m}\,Q_\mathrm{SQM}=\left(\frac{0.1}{0.7}\right)^{9/2}
  \dot{m}_\mathrm{SB}\,Q_\mathrm{SQM} T_\mathrm{S,6}^9+\sigma T_\mathrm{S}^4\;.\label{combined}
\end{equation}
for a non-isothermal core. Here $\dot{m}$ is the time averaged
accretion rate in units of nucleons
cm$^{-2}$ s$^{-1}$ and $\dot{m}_\mathrm{SB}$ is the accretion rate for
superbursters.
\section{Consistency with observations}
\begin{figure*}[!ht]
\centering
\centerline{\psfig{file=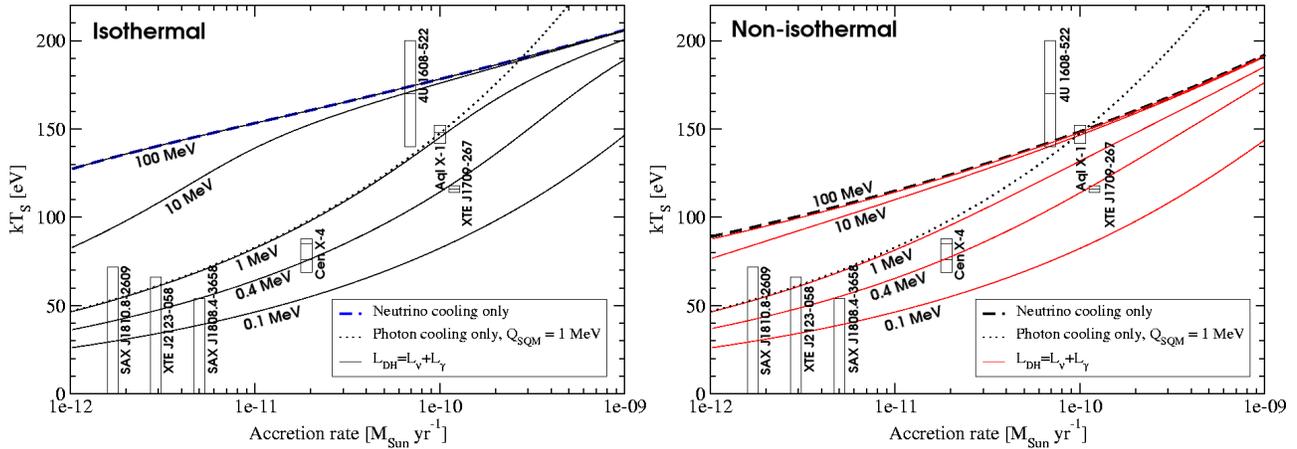,width=17cm,clip=} }
\caption{{\it Left}: comparing combined photon and neutrino cooling of
  an isothermal core (Eq.~\ref{isocombined} with
  $\dot{m}_\mathrm{SB}=0.3\,\dot{m}_\mathrm{Edd}$) to observations
  with $Q_\mathrm{SQM}$ as labelled for each curve. For reference
  curves with neutrino cooling only (Eq.~\ref{iso}) and photon cooling
  only are also shown. The curves assuming $Q_\mathrm{SQM}=100$ MeV
  and neutrino cooling only are nearly identical. For Cen X-4 the
  quoted temperature intervals overlap. {\it Right}: corresponding
  curves for a non-isothermal core.}
\label{fig1}
\end{figure*}
Fig. \ref{fig1} compares the temperatures derived above to observed
soft X-ray transient surface temperatures including pure blackbody
cooling, pure neutrino cooling from Eqs. \ref{iso} and \ref{niso} and
the combined case in Eqs. \ref{isocombined} and \ref{combined}. The
observational sample is discussed in \cite{Stejner:2006} and consists
of soft X-ray transients for which X-ray observations have determined
both the surface temperature and accretion rate. The accretion rate is
very uncertain however, so only thick bars are plotted, and for the three
coldest sources only upper limits to the temperature have been
established.

As may be seen in Fig. \ref{fig1} photon cooling dominates at low
accretion rates and for low binding energies, $Q_\mathrm{SQM}$. For
warmer sources with high accretion rates or if the binding energy is
high neutrino cooling dominates. No single curve can fit all the
sources, but the heating mechanism discussed here, although
unavoidable, may not be the only relevant source of heat -- indeed the
powerlaw components in the spectra of some of these sources are often
taken to indicate residual accretion in quiescence -- and so it only
provides a minimum luminosity. The surface temperatures should
therefore not fall significantly short of the of the temperature
curves shown in Fig. \ref{fig1} to be consistent with the core
parameters required for superburst ignition at high accretion
rates. Hence consistency of this model requires a very low binding
energy below 1 MeV per nucleon.

It may be noted that Brown et al.(1998) similarly found that deep
crustal heating of neutron stars could explain the quiescent emission
from Cen X-4 if only 0.1 MeV of the energy released by pycnonuclear
reactions per nucleon during an outburst was conducted into the core
and deposited there.~\cite{Yakovlev:2003,Yakovlev:2004}~instead
interpreted Cen X-4 and SAX J1808.4-3658 as massive neutron stars with
enhanced neutrino emission. In the context of strange stars the energy
is deposited directly in the core, but no enhanced neutrino cooling is
required if the quark matter binding energy, $Q_\mathrm{SQM}$, is small.

\section{Time dependence of the heating mechanism}
The model presented above used the time averaged accretion rate, but
the accretion rate for soft X-ray transients is of course extremely
time dependent, and if the transfer of matter from crust to core
mirrors this time dependence such an approximation may not be adequate
for all purposes. Modelling this in full here would go to far and only
the time dependence of the heating mechanism (i.e. the tunneling rate
of ions in the crust through the Coulomb barrier at the bottom of the
crust) will be discussed, but one may be guided with respect to its
consequences by similar considerations for neutron stars.
 
In neutron stars \cite{Ushomirsky:2001} found that following an
accretion outburst the heat released by pycnonuclear reactions and
electron captures higher in the crust forms distinct heat waves
reaching the surface by diffusion. By analogy if the tunneling rate at
the bottom of the crust -- and hence the heating -- is sufficiently
time dependent following an accretion event, one would expect the
qualitative conclusions in \cite{Ushomirsky:2001} to apply, leading to
strong variations in the surface temperature following an outburst as
the resulting heat waves reach the surface. In strange stars the heat
is released at a lower density in the crust with a shorter diffusion
time to the surface however, so the two heat waves may be closer in
time.

Transfer of matter from the crust, with mass $M_\mathrm{C}$, to the core
by tunneling takes place at a rate of 
(Stejner \& Madsen 2005)
%\cite{Stejner:2005}

\begin{align}
\frac{\mathrm{d}M_\mathrm{C}}{\mathrm{d}t} 
=-23.8\frac{\rho_\mathrm{b}}{\rho_\mathrm{D}}\tau(\rho_\mathrm{b}) \mbox{ M}_\odot \mbox{ s}^{-1}\label{dMdt}
\end{align}
where $\rho_\mathrm{b}$ is the density at the bottom of the crust,
$\rho_\mathrm{D} =7.8 \times 10^{11}\mbox{ g cm}^{-3}$ is the neutron
drip density and $\tau$ is the transmission coefficient for ions
striking the Coulomb barrier in their lattice motion. In
\cite{Stejner:2005} we discussed the structure of the gap between
crust and core and found an expression for the transmission
coefficient, which when inserted in Eq. \ref{dMdt} allows us to solve
for crust mass and tunneling rate as functions of time for a given
accretion scenario.
\begin{figure}[!htbp]
\centering
\includegraphics[width=8.5cm]{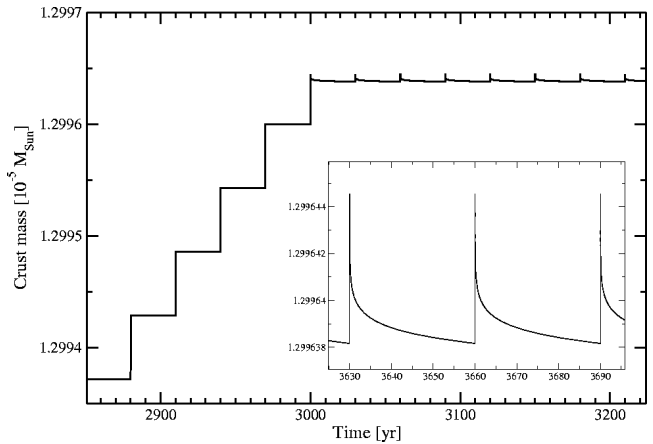}
\caption{A crude model for crust mass variations in Cen X-4. The inset
  shows a few equilibrium cycles.}
\label{fig2}
\end{figure}
One such solution is shown in Fig. \ref{fig2} for a system with 10 day
long accretion outbursts separated by quiescent intervals of 30 years
and an average accretion rate of $1.9\times 10^{-11} \mbox{ M}_\odot
\mbox{ yr}^{-1}$ -- roughly corresponding to \object{Cen X-4}. The
integration is started with a crust mass sufficiently low that the
tunneling rate is negligible and the crust builds up until the mass
transferred to the core during each cycle matches the average
accretion rate. At this point the crust mass is at its maximum -- as
determined by the choice of Coulomb barrier height -- during the
accretion outbursts and then relaxes to a state with very little
tunneling within about a year after each outburst. This is because the
tunneling rate increases sharply with density near equilibrium, and
means that practically all the heating will take place during or
immediately after an accretion outburst. Hence there will be heat
waves similar to those found by \cite{Ushomirsky:2001} following an
outburst, and these might be detectable in systems such as Cen X-4
where the recurrence time is very long (no outburst since 1969)
allowing them to reach the surface before they are smeared out by the
next outburst.

If -- speculatively -- strange stars are able to cool significantly on
timescales comparable to the recurrence times for soft X-ray
transients, it is also possible that the temperature will follow the
tunneling rate more closely, and that the current tunneling rate late
in an accretion cycle is therefore a better predictor for the
temperature than the time averaged rate. For the solution in
Fig. \ref{fig2} and using Eq. \ref{niso} (neutrino cooling only) this
actually predicts a surface temperature of 76 eV late in the cycle --
just as observed for Cen X-4. To demonstrate the effect credibly
would require detailed modelling beyond our scope here, but this does
show the potential consequences: the need to keep $Q_\text{SQM}$ small
to explain such sources is reduced to the point that even neutrino
emission alone could explain cold sources with long recurrence
times. The coldest of them all, SAX J1808.4-3658, has a recurrence
time of only 2 years however, so there the average accretion rate
should be adequate casting doubt on the relevancy of such an argument. 
 
\section{Discussion}
Using a simple scaling argument for the steady state temperature of
accreting strange stars a consistency check was presented between
the constraints on strange quark matter derived from superburst
ignition conditions, and the temperature of quiescent soft X-ray
transients -- both assumed to derive from deep crustal heating by
transfer of matter from crust to core. 

We have seen that although the hottest soft X-ray transients most similar
to superbursters are always consistent with superbursters, the colder
sources require very low binding energy for strange quark matter below
an MeV, unless the time dependence is more pronounced than would
ordinarily be expected. This seems conspicuously fine tuned given that
we are working with strong interaction energy scales, but no clear
inconsistency could be shown. 

\begin{figure}[!htbp]
\centering
\includegraphics[width=8.5cm]{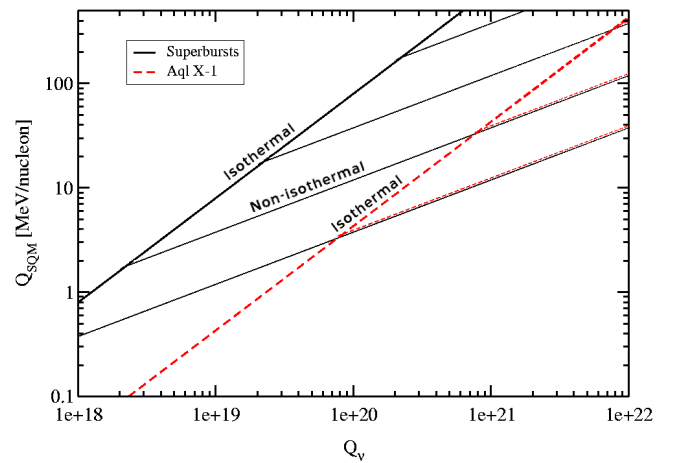}
\caption{Constraints on quark core parameters obtained from
  superbursters and the soft X-ray transient Aql X-1. Thin lines show
  non-isothermal constraints from Eq.~\eqref{niso} with $K=10^{19},\,
  10^{20},\, 10^{21},\, 10^{22}\mbox{ cgs}$ from the bottom
  up. Non-isothermal constraints for \object{Aql X-1} are nearly
  indistinguishable from those for superbursters. Thick lines show
  isothermal constraints from Eq.~\eqref{iso}. Here $\dot{m}=0.3
  \dot{m}_\mathrm{Edd}$ for the superbursters.}
\label{fig3}
\end{figure}

In principle this simple model can go no further than this consistency
check, but to demonstrate the method Fig.~\ref{fig3} shows how one
might proceed to further constrain the quark matter
properties. It shows the relations between the core parameters, which from
Eqs. \ref{iso} and \ref{niso} must be fulfilled to reach superburst
ignition and to explain Aql X-1 with neutrino cooling
alone. Increasing the thermal conductivity one reaches
$K_\mathrm{crit}$ at which point the core must be isothermal, so there
are no common solutions between the isothermal curves. Hence the
presence of an additional heating source for Aql X-1 must be assumed
to render the constraints for this source irrelevant and allow
solutions with low $Q_\mathrm{SQM}$ consistent with the colder
sources. However even then we see from the superburst constraints that
we would be limited to very low values of $K$ and $Q_\nu$ as
well. Although the numerical values of these constraints could change
significantly with more detailed modelling this shows the potential
for such an approach to limit the possible range of core parameters
and ultimately the range of QCD-parameters consistent with the strange
quark matter hypothesis.
\acknowledgement{Some of the work reported
here was performed in collaboration with Jes Madsen.}

%\end{document}

          \clearpage


\begin{thebibliography}{13}
\expandafter\ifx\csname natexlab\endcsname\relax\def\natexlab#1{#1}\fi

\bibitem[{Alcock} {et~al.}(1986)]
%{Alcock}, {Farhi}, \& {Olinto}}]
{Alcock:1986} {Alcock}, C., {Farhi}, E., \& {Olinto}, A. 1986, ApJ, 310, 261

\bibitem[{Brown} {et~al.}1998]
%{Brown}, {Bildsten}, \& {Rutledge}}]
{Brown:1998}
{Brown}, E.~F., {Bildsten}, L., \& {Rutledge}, R.~E. 1998, ApJ, 504, L95

\bibitem[{Cumming} {et~al.}(2005)]
%{Cumming}, {Macbeth}, {in 't Zand}, \& {Page}}]
{Cumming:2005}
{Cumming}, A., {Macbeth}, J., {in 't Zand}, J.~J.~M., \& {Page}, D. 2005,
  [arXiv:astro-ph/0508432]

\bibitem[{Gudmundsson} {et~al.}(1983)]
%{Gudmundsson}, {Pethick}, \& {Epstein}}]
{Gudmundsson:1983}
{Gudmundsson}, E.~H., {Pethick}, C.~J., \& {Epstein}, R.~I. 1983, ApJ, 272,
  286

\bibitem[{Haensel} \& {Zdunik} 1990]{Haensel:1990}
{Haensel}, P. \& {Zdunik}, J.~L. 1990, A\&A, 229, 117

\bibitem[{Haensel} \& {Zdunik} 2003]{Haensel:2003}
{Haensel}, P. \& {Zdunik}, J.~L. 2003, A\&A, 404, L33

\bibitem[{Kuulkers}(2004)]{Kuulkers:2004}
{Kuulkers}, E. 2004, Nuclear Physics B Proceedings Supplements, 132, 466

\bibitem[{Page} \& {Cumming}(2005)]{Page:2005}
{Page}, D. \& {Cumming}, A. 2005, ApJ, 635, L157

\bibitem[{Stejner} \& {Madsen}(2005)]{Stejner:2005}
{Stejner}, M. \& {Madsen}, J. 2005, Phys. Rev. D, 72, 123005

\bibitem[{Stejner \& Madsen(2006)}]{Stejner:2006}
Stejner, M. \& Madsen, J. 2006, [ArXiv:astro-ph/0603566]

\bibitem[{Ushomirsky} \& {Rutledge}(2001)]{Ushomirsky:2001}
{Ushomirsky}, G. \& {Rutledge}, R.~E. 2001, MNRAS, 325, 1157

\bibitem[{Yakovlev} {et~al.}(2003)]
%{Yakovlev}, {Levenfish}, \& {Haensel}}]
{Yakovlev:2003}
{Yakovlev}, D.~G., {Levenfish}, K.~P., \& {Haensel}, P. 2003, ApJ, 407, 265

\bibitem[{Yakovlev} {et~al.}(2004)]
%{Yakovlev}, {Levenfish}, {Potekhin}, {Gnedin}, \& {Chabrier}}]
{Yakovlev:2004}
{Yakovlev}, D.~G., {Levenfish}, K.~P., {Potekhin}, A.~Y., {Gnedin}, O.~Y., \&
  {Chabrier}, G. 2004, ApJ, 417, 169

\end{thebibliography}
\end{document}